\documentclass[a4paper,11pt]{article}
\usepackage{graphicx}
\usepackage{epsfig}
\usepackage{color}
\usepackage{colordvi}

\newcommand{\BABARPubYear}    {06}

\newcommand{\BABARConfNumber} {003}
\newcommand{\SLACPubNumber} {11971}

\input pubboard/babarsym

\long\def\inst#1{\par\nobreak\kern 4pt\nobreak
    {\it #1}\par\vskip 10pt plus 3pt minus 3pt}

\begin{document}
{\pagestyle{empty}


\begin{flushright}
\babar-CONF-\BABARPubYear/\BABARConfNumber \\
SLAC-PUB-\SLACPubNumber \\
July 2006 \\
\end{flushright}

\par\vskip 5cm

\begin{center}
\Large \bf  \mathversion{bold}
Search for Charmonium States Decaying to $ J/\psi \gamma \gamma $
Using
Initial-State Radiation Events
\end{center}
\bigskip

\begin{center}
\large The \babar\ Collaboration\\
\mbox{ }\\
\today
\end{center}
\bigskip \bigskip

\begin{center}
\large \bf Abstract
\end{center}
We study the processes
$e^+e^-\to  ( J/\psi \gamma \gamma ) \gamma $
and
$e^+e^-\to (J/\psi \pi^- \pi^+) \gamma $
where the hard
photon radiated from an initial $ e^+ e^- $ collision
with center-of-mass (CM) energy near 10.58 GeV is detected.
In the final state  $  J/\psi \gamma \gamma  $
we consider $ J/\psi \pi^0 $,
$  J/\psi \eta $, $ \chi_{c1} \gamma $, and
 $ \chi_{c2} \gamma $ candidates.
The
invariant mass of the hadronic final state defines the effective \epem
CM energy in each event, so these
data can be compared with
direct \epem measurements.
We report 
90\% CL upper limits for the
integrated cross section times branching fractions
of the  $ J/\psi \gamma \gamma $
channels in the $ Y(4260) $ mass region.
\vfill
\begin{center}

Submitted to the 33$^{\rm rd}$ International Conference on High-Energy Physics, ICHEP 06,\\
26 July---2 August 2006, Moscow, Russia.

\end{center}

\vspace{1.0cm}
\begin{center}
{\em Stanford Linear Accelerator Center, Stanford University, 
Stanford, CA 94309} \\ \vspace{0.1cm}\hrule\vspace{0.1cm}
Work supported in part by Department of Energy contract DE-AC03-76SF00515.
\end{center}

\newpage
} 

\begin{center}
\small

The \babar\ Collaboration,
\bigskip

%
{B.~Aubert,}
{R.~Barate,}
{M.~Bona,}
{D.~Boutigny,}
{F.~Couderc,}
{Y.~Karyotakis,}
{J.~P.~Lees,}
{V.~Poireau,}
{V.~Tisserand,}
{A.~Zghiche}
\inst{Laboratoire de Physique des Particules, IN2P3/CNRS et Universit\'e de Savoie,
 F-74941 Annecy-Le-Vieux, France }
% ...

\end{center}\newpage


\section{INTRODUCTION}
\label{sec:Introduction}

In 2005, \babar\ reported the first
observation of a broad structure
in the $ J/\psi \pi^- \pi^+ $ mass spectrum produced via
$ e^+ e^- $ annihilation and detected in
initial state radiation (ISR) events:
$e^+e^-\to (J/\psi \pi^- \pi^+) \gamma $.
The mass was reported to be approximately $ 4.26 \ \mbox{GeV}/c^2 $
and the width $ \approx 90 \ \mbox{MeV}/c^2 $ if interpreted
as a single resonance \cite{shuwei}.
Subsequently, CLEO-c reported cross section 
measurements for $ J/\psi \pi^- \pi^+ $ and $ J/\psi \pi^0 \pi^0 $
production at this mass, confirming the large production rate
\cite{ref:CleoY4260Scan}.
More recently, they have reported ISR measurements
of the $ J/\psi \pi^- \pi^+ $ final state \cite{ref:CLEOFPCP2006}.
As discussed in refs. \cite{shuwei}, \cite{ref:CleoY4260Scan},
\cite{ref:CLEOFPCP2006}, and references therein,
the nature of the $ Y(4260) $ is not clear, so
observing additional decay modes and measuring their
relative rates is important for advancing our understanding of
this state and its production rate.

In this paper we report measurements of
$  e^+ e^- \to J/\psi \gamma \gamma  $ cross sections
using ISR events 
where the hard
photon radiated from an initial $ e^+ e^- $ collision
is detected directly.
Requiring that the ISR photon is detected reduces
backgrounds significantly and increases the
signal-to-background ratio,
which we confirm with a parallel study of 
the benchmark signal $ e^+ e^- \to J/\psi \pi^- \pi^+ $.
In the final state  $  J/\psi \gamma \gamma $
we consider $ J/\psi \pi^0 $,
$  J/\psi \eta $, $ \chi_{c1} \gamma $, and
 $ \chi_{c2} \gamma $ candidates.
If the $ Y (4260) $ is a charmonium state,
its decays  might be
similar to those of the $ \psi({\rm 2S}) $
which has relatively large branching fractions to
$  J/\psi \eta $, $ \chi_{c1} \gamma $, and
 $ \chi_{c2} \gamma $ \cite{ref:pdg2004}.

The ISR cross section for a particular hadronic final state $f$ (excluding
the radiated photon) is related to the
corresponding \epem cross section $\sigma_f(s)$ by:
\begin{equation}
\frac{d\sigma_f(s,x)}{dx} = W(s,x)\cdot \sigma_f(s(1-x))\ ,
\label{eq1}
\end{equation}
where $x=2E_{\gamma}/\sqrt{s}$; $E_{\gamma}$ is the
energy of the ISR photon in the nominal \epem  center-of-mass (CM)\@ frame; 
$\sqrt{s}$ is
the nominal \epem CM\@ energy; and $\sqrt{s(1-x)}$ is the effective
CM\@ energy
at which the final state $f$ is produced. The
invariant mass of the hadronic final state defines the effective \epem
CM energy.
The function $W(s,x)$ is calculated with better than 1\% accuracy \cite{ivanch}
and describes the
probability density function for ISR photon emission,
which occurs at all angles. 
For
the present study we require that the hard ISR photon be detected in the
electromagnetic calorimeter of the \babar\ detector.
The effective ISR luminosities at CM energies corresponding to
the masses of the $ J/\psi $, the $ \psi(2S) $,
and the $ Y(4260) $ are
 66.3 MeV${}^{-1}$ nb${}^{-1}$,
 84.3 MeV${}^{-1}$ nb${}^{-1}$,
and
 105 MeV${}^{-1}$ nb${}^{-1}$.

\section{THE \babar\ DETECTOR AND DATASET}
\label{sec:babar}

The data used in this analysis were collected with the \babar\ detector
at the \pep2\ storage ring. 
The data set used here is the same data set studied
in the original $ Y(4260) $ analysis \cite{shuwei};
the integrated luminosity at and near 10.58 GeV
in the  $ e^+ e^- $ CM is approximately $ 230 $
fb${}^{-1}$.

Charged-particle momenta are measured in a tracking
system consisting of a five-layer double-sided silicon
vertex tracker (SVT) and a 40-layer central drift chamber
(DCH), both situated in a 1.5-T axial magnetic field.
An internally reflecting ring-imaging Cherenkov detector
(DIRC) with bar radiators made of fused silica
provides charged-particle
identification.
A CsI electromagnetic calorimeter (EMC) is used to detect
and identify photons and electrons.
Muons are identified in the instrumented magnetic
flux return system (IFR).

Electron candidates are identified by the ratio of
the shower energy deposited in the EMC to the
momentum, the shower shape, the specific ionization
in the DCH, and the Cherenkov angle measured by the
DIRC.
Muons are identified by the depth of penetration
into the IFR, the IFR cluster geometry, and the
energy deposited in the EMC.
Pion candidates are selected based on a likelihood
calculated from the specific ionization in the
DCH and SVT, and the Cherenkov angle measured in the
DIRC.
Photon candidates are identified with clusters
in the EMC that have a shape consistent with an
electromagnetic shower but without an associated
charged track.
A  detailed
description of the \babar\ detector is available
elsewhere~\cite{ref:babar}.

We use an
admixture of simulation packages 
for the final states studied.
These packages use adaptations of
algorithms for
generating hadronic final
states via initial state radiation \cite{ref:CzyzKuhn2001},
for modeling
multiple soft-photons from initial-state 
electrons and positrons via
a structure-function technique  
\cite{ref:Arbuzov1997, ref:Caffo1997}, 
and 
for modeling photons from
final-state particles \cite{ref:Photos}. 
We pass the generated events through a
detector simulation based on GEANT4 \cite{ref:GEANT4},
and reconstruct them in the same
way as we do the data.

\section{EVENT SELECTION CRITERIA}
\label{sec:Analysis}

The event selection criteria used in this study, and described below, 
were chosen to produce very low backgrounds and
to eliminate events with relatively poor
mass resolution. 
These goals are especially important for the
high-mass  $ J/\psi \gamma \gamma $ final state where
the anticipated signal level is small.
We therefore require that the ISR photon be detected directly 
for this analysis and do not use the much larger sample of
events where the ISR photon is produced outside the 
EMC acceptance.
To avoid bias, possible signals in the mass region above 3.85 GeV/$c^2$,
were not studied until all of the selection criteria
had been determined.

All charged tracks
used in this analysis
are required to
have laboratory polar angle in the
range 0.45 to 2.40 radians and are required to have at least
one hit in the SVT and at least 20 hits in the DCH.
Such tracks are reconstructed relatively well and
are detected in a fiducial region where
the particle identification
is understood especially well.
Candidate $ J/\psi $ mesons for the $ J/\psi \gamma \gamma $ analysis
are identified via their decays to  $ \mu^+ \mu^- $.
Those for the 
$ J/\psi \pi^- \pi^+ $ analysis are identified
via their decays to $ \mu^+ \mu^- $ and
$ e^+ e^- $.
We require that both tracks in a dimuon candidate be 
identified as muons, with at least
one satisfying a set of criteria
which may be referred to as loose and the
other satisfying a set of criteria 
which may be referred to as
very loose.
Using a sample of $ \psi({\rm 2S}) \to J/\psi \pi^- \pi^+ $
decays produced in ISR events,
we measure the muon identification efficiency for pairs
to be
$ \approx 80 \% $.
Because QED production of four-lepton events
is a major source of background for the
$ J/\psi \pi^- \pi^+ $ sample, we exclude
all candidates where either of the pion
candidates is identified as possibly a muon 
or an electron.
To improve the signal-to-background ratio, and
because the $ J/\psi \to e^+ e^- $ signal has a long
radiative tail and thus spans a greater mass range than
does the $ J/\psi \to \mu^+ \mu^- $ signal,
we require that both charged tracks be positively
identified as electrons in selecting $ e^+ e^- $
pairs.
We do not measure the electron pair efficiency {\em
per se}, but measure the $ J/\psi \to e^+ e^- $
efficiency relative to the $ J/\psi \to \mu^+ \mu^- $
efficiency using ISR $ \psi({\rm 2S}) \to J/\psi \pi^- \pi^+ $
events.

ISR events are characterized by small missing mass,
corresponding to that of the ISR photon,
recoiling from the exclusive 
hadronic final state of interest.
Experimentally, the missing mass squared, measured
using the beam momentum and the 
momenta of the final state hadrons
(including the non-ISR photons), is not exactly zero.
Most importantly, higher order QED contributions
to ISR production of the final state produce
a very long tail at positive missing mass.
In addition,
the finite spread of the beam energy/momentum and the
the detector resolution  further distort the
observed values.
However, the signal is predominantly produced
with small recoil mass, and the signal-to-background is
greatest when the observed recoil mass squared is close
to zero.
We exploit this feature of the process both to
select a sample of candidates with lower background
rate and to improve the resolution of events selected.
We require the recoil mass squared, determined from the
final state particles' measured momenta and the beam particles'
nominal momenta, to lie in the range
-2.0 to 4.0 GeV$^{2}/c^{4}$.
We also require that the missing energy,
calculated in the lab, fall in the range
-0.4 GeV to 1.5 GeV and that no more than 400 MeV
of additional neutral energy be detected in the EMC.
We then re-fit the hadronic final state particles' momenta
using the nominal 
beam particles' momenta,
the relevant track parameter error matrices,
and
the constraint
that the recoil mass
is zero.
We call this the 1C fit
and require its $ \chi^2 $ to be 
less than 20.
The 1C fit predicts a direction for the ISR photon.
We require that the direction of the observed ISR
photon candidate coincide with this predicted
direction within 15 mr.
In addition to reducing backgrounds,
this last requirement, in the absense of any others,
rejects approximately 25\% of all signal candidates
as a result of the higher order QED processes that
contribute to the ISR cross section.
In such events, the 1C fit $ \chi^2 $ must also be
high, so the pointing and $ \chi^2 $ criteria are
strongly correlated.

We select as $ J/\psi $ candidates dimuon pairs
whose mass lies within 20 MeV/$c^2$ of the nominal
$ J/\psi $ mass \cite{ref:pdg2004}.
The 1C fit improves the dimuon resolution significantly, but
as our charged track resolution is better than
our photon resolution, this effect is less pronounced
for the $ J/\psi \gamma \gamma $ sample
(where the $ J/\psi $ mass resolution is $ \approx 12 $~MeV/$c^2$)
than for the $ J/\psi \pi^-  \pi^+ $  sample
(where the $ J/\psi $ mass resolution is $ \approx 7 $~MeV/$c^2$).
However, as the signal-to-background (in the sample
of $ J/\psi $ candidates) is worse 
in the $ J/\psi \gamma \gamma $ sample, we use
the same dimuon mass criteria for both $ J/\psi \pi^- \pi^+ $
and  $ J/\psi \gamma \gamma $  final states.
For the $ J/\psi \pi^- \pi^+ $ channel we also 
select as $ J/\psi $ candidates
electron pairs whose mass lies in the range 3.050
- 3.120 GeV/$c^2$.
This asymmetric window around the nominal $ J/\psi $
mass is used because the signal has a long radiative tail.
The specific limits of the window were chosen following
studies of 
ISR production of
$ \psi({\rm 2S} ) \to J/\psi \pi^- \pi^+ $
with $ J/\psi \to e^+ e^- $.
We do not use $ J/\psi \to e^+ e^- $ candidates in
this study of $ J/\psi \gamma \gamma $  final states
as the ISR data set used in this analysis was defined before the
discovery of the $ Y(4260) $,
and excluded $ e^+ e^- \gamma \gamma \gamma $
candidates intentionally.

For the $ J/\psi \gamma \gamma $
candidates accepted by the selection criteria
described above, a two-constraint (2C) fit
is done where the invariant mass of the
lepton pair is additionally constrained to
be that of the $ J/\psi $.
We then explicitly require the presence of intermediate
states so that $ J/\psi \gamma \gamma $ is a manifestation of 
$  J/\psi \eta $, 
$  J/\psi \pi^0 $, $ \chi_{c1} \gamma $, or
 $ \chi_{c2} \gamma $.
We require that each of the non-ISR photons has
measured energy in the laboratory of at least 100 MeV
and that its laboratory polar angle lies in the
range 0.35 to 2.40 radians, where we understand
the EMC performance well.
Using momenta from the 2C fit,
we require  that the invariant mass of a photon
pair lies in the range 0.120 - 0.150 MeV/$ c^2 $
to be considered as a $ \pi^0 $ candidate and in
the range 0.52 - 0.58  MeV/$ c^2 $
to be considered as an $ \eta $ candidate.
Similarly, we require that a $ J/\psi \gamma $
invariant mass lies in the range 3.4906 - 3.5309
GeV/$ c^2 $ (3.5363 -  3.5763 GeV/$ c^2 $) 
to be considered as a $ \chi_{c1} $ ($ \chi_{c2} $)
candidate.
Events with such candidates are then re-fit with
a third constraint requiring that the invariant
mass of the corresponding combination be that
of the hypothesis.
When an event satisfies the nominal requirements
for more than one hypothesis, the hypothesis with
the best 3C $ \chi^2 $ is selected as that to be
used in the final analysis.
Given our  $ \gamma \gamma $ and $ J/\psi \gamma $ resolutions,
these windows accept about 90\% of the candidates
otherwise accepted.
The $ J/\psi \gamma \gamma $ resolutions depend
upon the mass of the final state, in particular how
far above threshold it is.
For the decay $ Y(4260) \to J/\psi \eta $,
the (simulated) resolution is $ \approx 10 $~MeV/$c^2$.
For all of the $ Y(4260) $ decay modes considered,
the mass resolution is negligible relative to its
$ \approx 90 $~MeV/$c^2$ width.

\begin{figure}[htb]
\begin{minipage}{2.9in}
\centering
\mbox{\epsfxsize7.5cm\epsffile{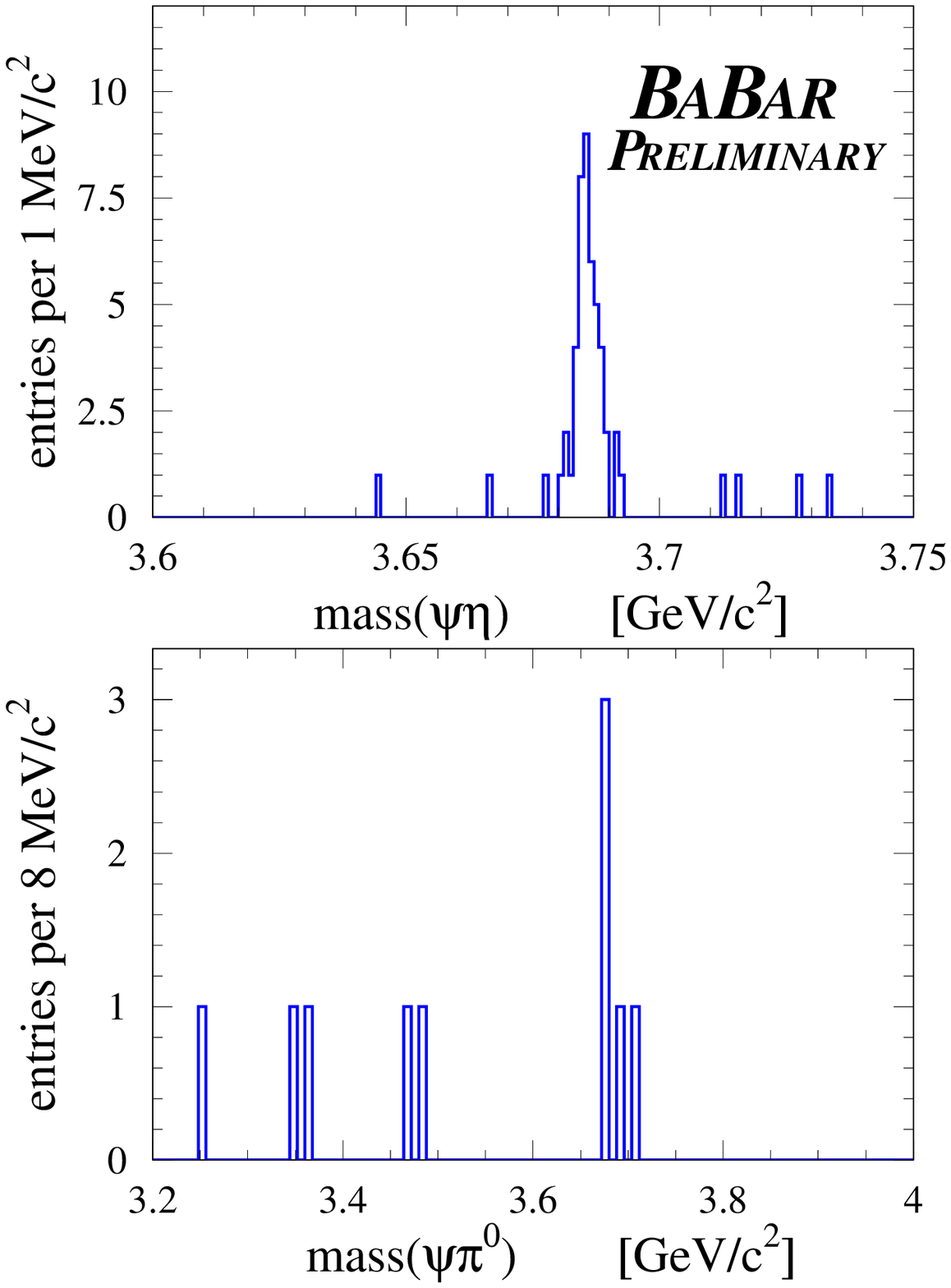}}
\caption{
The upper plot shows
the $ J/\psi \eta $ invariant mass 
for $ J/\psi \gamma \gamma $
candidates where $ J/\psi \eta $ is the best hypothesis.
The lower plot shows the  $ \psi \pi^0 $ invariant mass 
for $ J/\psi \gamma \gamma $
candidates where $ J/\psi \pi^0 $ is the best hypothesis.
} 
\label{PsiGG1}
\end{minipage}
\quad
\begin{minipage}{2.9in}
\mbox{\epsfxsize7.5cm\epsffile{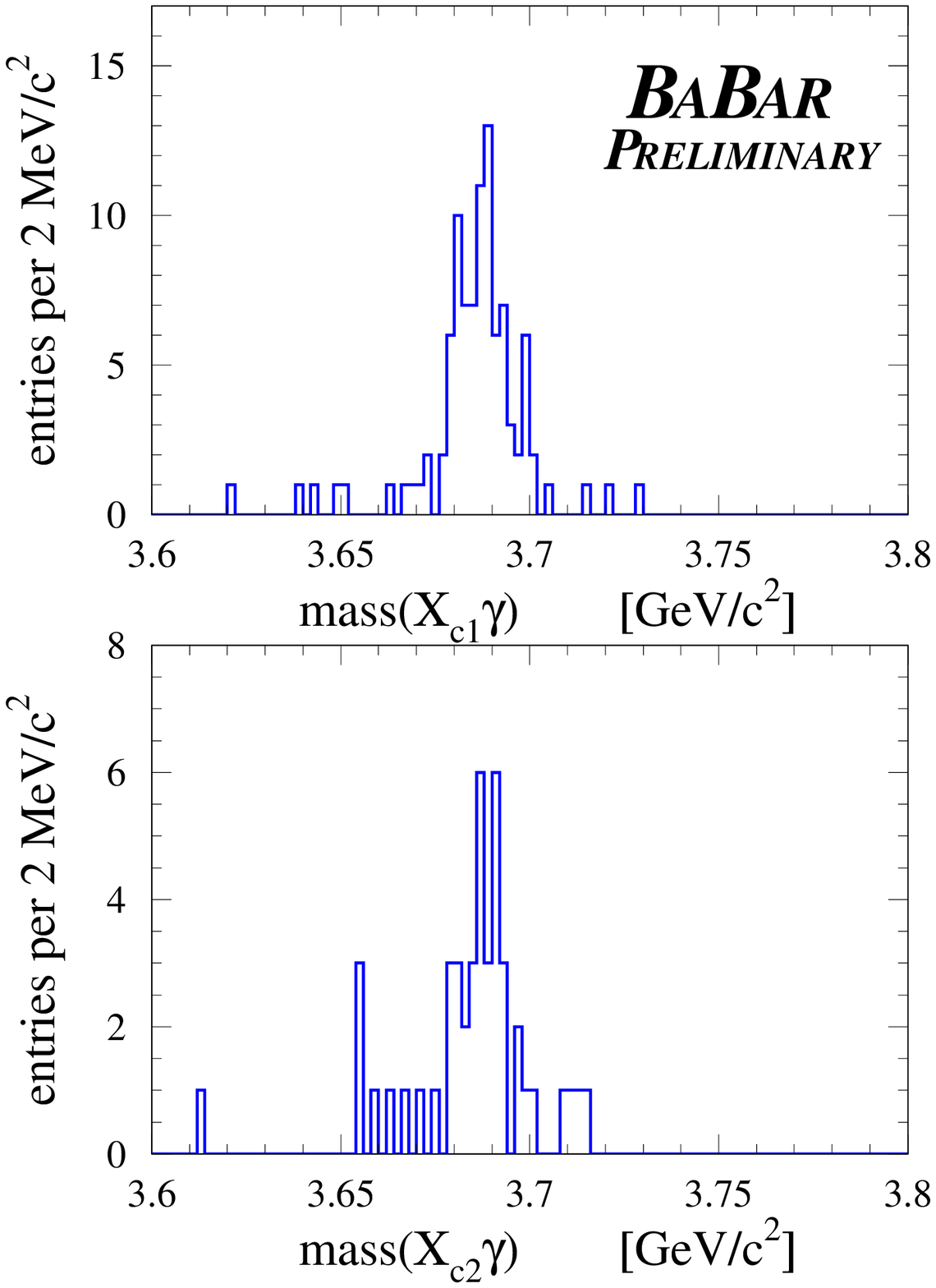}}
\caption{
The upper plot shows
the $ \chi_{c1}\, \gamma $ invariant mass 
for $ J/\psi \gamma \gamma $
candidates where $  \chi_{c1}\, \gamma $ is the best hypothesis.
The lower plot shows the  $  \chi_{c2}\, \gamma  $ invariant mass
for $ J/\psi \gamma \gamma $
candidates where $  \chi_{c2}\, \gamma $ is the best hypothesis.
} 
\label{PsiGG2}
\end{minipage}
\end{figure}

\section{\mathversion{bold}MEASUREMENTS AT THE $ \psi({\rm 2S}) $}

We initially look at signals in the $ \psi({\rm 2S}) $
mass region but not above 3.85 GeV/$c^2$.
We also estimate background levels by looking at candidates
with dimuon invariant mass in
$ J/\psi $ sidebands.
The $ \psi({\rm 2S}) $ mass region
 $ J/\psi \gamma \gamma $ invariant mass distributions
for candidates
selected as 
$  J/\psi \eta $ or
$  J/\psi \pi^0 $ are shown in Fig.~\ref{PsiGG1}.
The corresponding distributions for those selected 
as $ \chi_{c1} \gamma $ or
 $ \chi_{c2} \gamma $
are shown in Fig.~\ref{PsiGG2}.
Note that the mass ranges of the samples
differ.
In particular, the $ J/\psi \pi^0 $ threshold is
much lower, and therefore its mass resolution is
much worse.

The signal levels are low in all four channels,
and the background levels much lower, so we estimate
the signal in each channel as the number of
entries within 15 MeV/$c^2$ of the nominal
$ \psi({\rm 2S}) $ mass.
This gives 46, 4, 84, and 34 events for the
$  J/\psi \eta $, 
$  J/\psi \pi^0 $, 
$ \chi_{c1} \gamma $, and
$ \chi_{c2} \gamma $ channels.
Monte Carlo simulations indicate that
the efficiencies for these channels are roughly
equal
(not including
the branching fractions of the intermediate
states).
The probability that an event generated as $ \chi_{c1} \gamma $
will be reconstructed as 
$ \chi_{c2} \gamma $ or as $ J/\psi \eta $
is of order 5\%.
The probabilities that an event generated
as $ J/\psi \eta $ or $ J/\psi \pi^0 $
will be reconstructed as $ \chi_{c1} \gamma $ or
$ \chi_{c2} \gamma $ 
are of order 5\% each.
Altogether, 
the number of 
$ \psi({\rm 2S} ) \to J/\psi \gamma \gamma $
signal events we observe is generally consistent
with, but somewhat
higher than,
the number projected from Monte Carlo studies using
branching fractions (or products of branching fractions)
from CLEO \cite{ref:CleoPsi2S}, which are generally higher than those
reported by the PDG \cite{ref:pdg2004} in 2004.

For both samples of $ J/\psi \pi^- \pi^+ $
candidates accepted by the selection criteria 
described above, a two-constraint (2C) fit
is done where the invariant mass of the
lepton pair is additionally constrained to
be that of the $ J/\psi $.
We observe  clean $ \psi({\rm 2S}) $ signals
with 785 candidates in the
$ J/\psi \to \mu^+ \mu^- $ sample
and 434 in the
$ J/\psi \to e^+ e^- $ sample.
In both channels the width of the peak, fit as a
Gaussian, is $ \approx 3 $ MeV/$c^2$ and the background
rate is at the percent level.
From the relative numbers of  $ \psi({\rm 2S}) $
signal events in the two $ J/\psi $ decay modes,
we find the combined acceptance times reconstruction
efficiency to be 1.55 times that for the
$ J/\psi \to \mu^+ \mu^- $ mode alone.
We assume the same ratio holds at higher
$ J/\psi \pi^- \pi^+ $ mass.

The integrated ISR luminosity, requiring the
ISR photon to be produced in the laboratory 
with a polar angle in the range 0.35 to
2.40 radians, where we understand the EMC
performance well, 
is 84.3 MeV${}^{-1}$ nb${}^{-1}$
at the mass of the
$ \psi({\rm 2S}) $,
and we estimate its fractional uncertainty to be 3\%.
For a narrow resonance, one may write the integrated cross section
\begin{equation}
\Sigma \equiv \int \sigma ( {\rm E_{CM}} ) d{\rm E_{CM}} 
\end{equation}
where $ \sigma $ is a resonant cross-section and is assumed
to have a Breit-Wigner line shape.
It is directly related to the partial decay rate to
electron pairs.
If the resonance has spin 1, then
\begin{equation}
\label{ResEqn}
\Sigma = {  { 6 \pi^2 } \over  {m^2} } \, \Gamma_{\rm ee}   \, .
\end{equation}
Using PDG values for $ \Gamma_{\rm ee} $,
$ m $, and relevant branching fractions \cite{ref:pdg2004},
we predict 784 
$ \psi({\rm 2S}) \to J/\psi \pi^- \pi^+; \ 
 J/\psi \to \mu^+ \mu^- $ events  with an 8\%
fractional error and we observe
785.
In a similar exercise, 
we predict $ \approx 17,900 $ ISR $ J/\psi \to
\mu^+ \mu^- $ events with a 4.6\% fractional uncertainty and
we observe $ 17,312 \pm 271 $ events, about 97\% of the predicted
number.
From these studies we conclude that we understand the
product of ISR luminosity and $  J/\psi \pi^- \pi^+ $
reconstruction efficiency with no worse than 10\%
precision.
\section{MEASUREMENTS AT HIGHER MASS}
\label{sec:Physics}

We estimate high-mass $ J/\psi \gamma \gamma $ background
rates due to non-$ J/\psi $ dimuon pairs
in the $ J/\psi $ mass range
using $ \mu^+ \mu^- $
background windows
75 to 175 MeV/$c^2$ above and below the nominal $ J/\psi $
mass.
For dimuon pairs in these windows we look for
$ \eta $ and $ \pi^0 $ candidates as we do for
real data (except for performing the 2C fit).
To create a sample of $ \chi_{C} $ candidates,
we calculate the $ \mu^- \mu^+ \gamma $ invariant
masses, subtract the $ \mu^- \mu^+ $ mass, and
add the $ J/\psi $ mass to compare to the
$ \chi_{c1} $ and $ \chi_{c2} $ masses.
In all four cases, we calculate an  effective
``$ J/\psi \gamma \gamma $"
mass as the $ \mu^- \mu^+ \gamma \gamma $ mass minus the
$ \mu^- \mu^+ $ mass plus the  $ J/\psi $ mass.
These background  windows enclose a total of 200 MeV/$c^2$
while the $ J/\psi $ candidates lie within a 40 MeV/$c^2$
wide window, so the background is estimated by dividing
the number of candidates observed by 5.
This predicts between 2 and 3 background events in
the $ J/\psi \eta $ sample, predominantly found below 5 GeV/$c^2$,
and about half this amount in each of the other channels.

\begin{figure}[thb]
\begin{minipage}{2.90in}
 \centering
\mbox{\epsfxsize7.5cm\epsffile{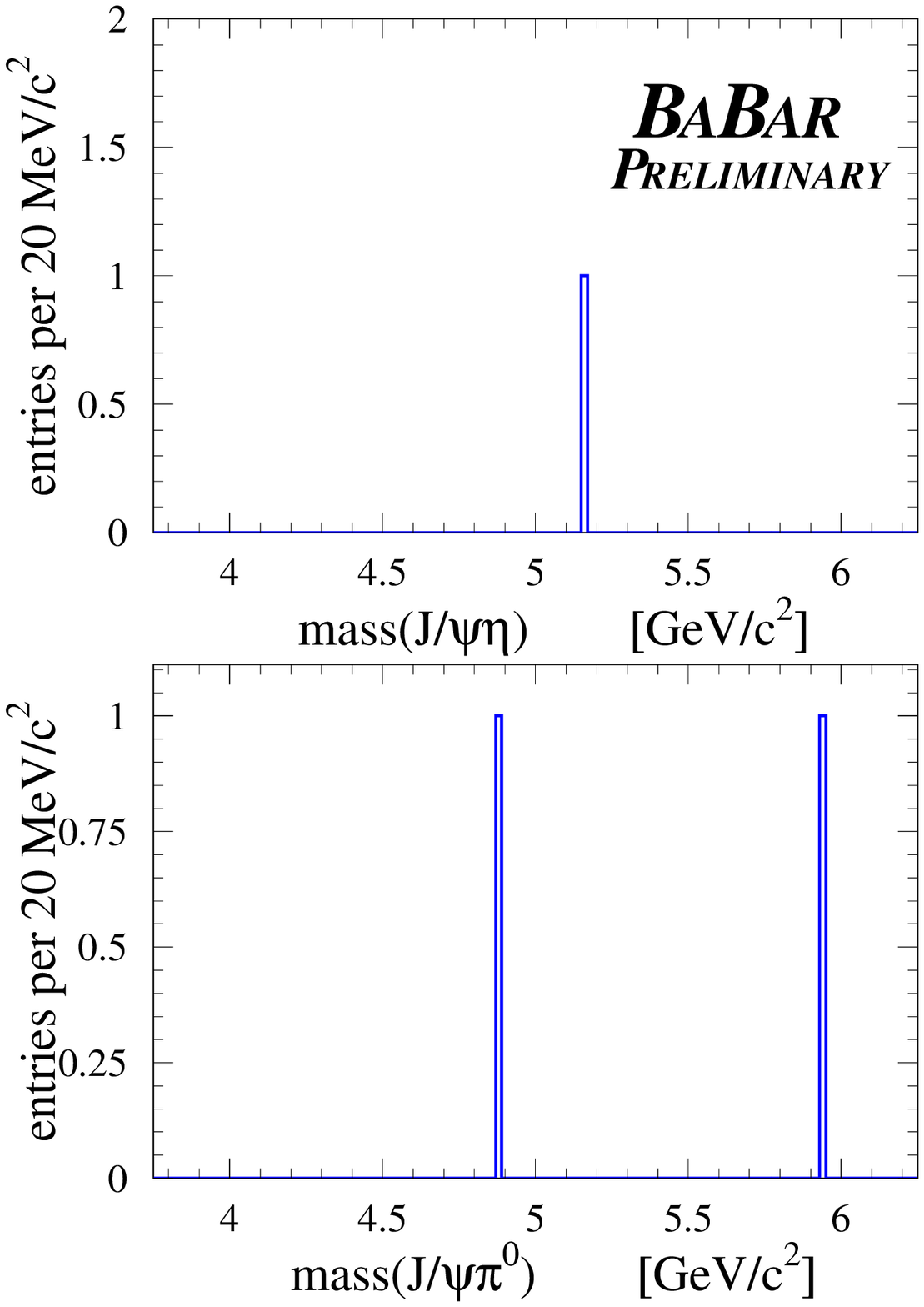}}
\caption{%
These are  the
$ J/\psi \, \eta $ (top) and
 $ J/\psi  \pi^0 $ (bottom)
invariant mass distributions
for candidates satisfying
all final analysis criteria.
}
\label{PsiGG5}
\end{minipage}
\quad
\begin{minipage}{2.90in}
 \centering
\mbox{\epsfxsize7.5cm\epsffile{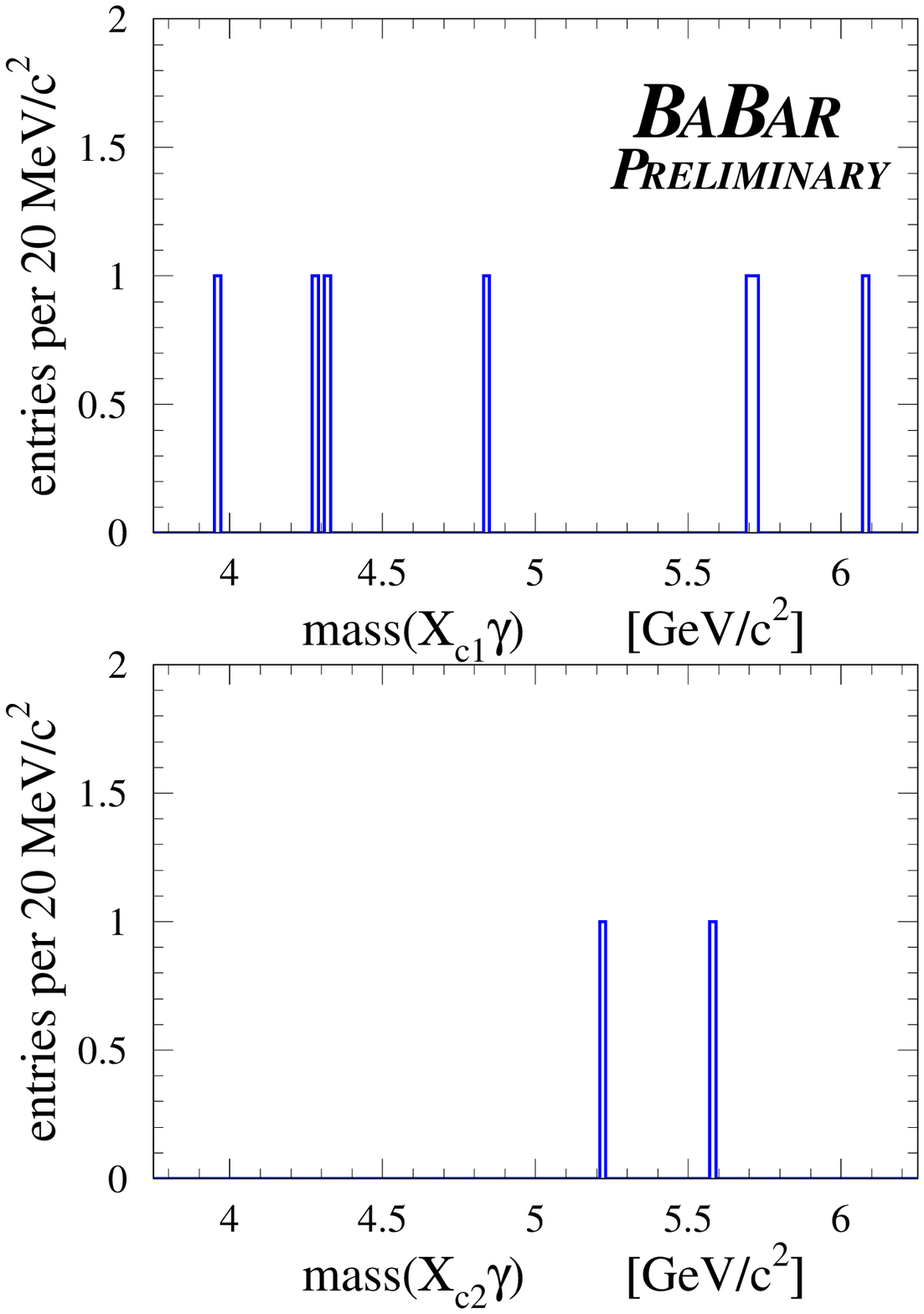}}
\caption{%
These are  the
$ \chi_{c1} \, \gamma $ (top) and
 $ \chi_{c2}\, \gamma  $ (bottom)
invariant mass distributions
for candidates satisfying
all final analysis criteria.
}
\label{PsiGG6}
\end{minipage}
\end{figure}

The  high mass $ J/\psi \eta $ and $ J/\psi \pi^0 $
data are shown in Fig.~\ref{PsiGG5}.
The  high mass $ \chi_{c1} \, \gamma $ and
$ \chi_{c2} \, \gamma $ data are shown in
Fig.~\ref{PsiGG6}.
  We see no events in the $ Y(4260) $ region in the $ J/\psi \, \eta $,
  $ J/\psi  \pi^0 $, or $ \chi_{c2}\, \gamma  $ distributions,
  and about the total number of entries estimated as
  non-$ J/\psi $ $ \mu^+ \mu^-  \gamma \gamma $.
  Two entries are observed in the $ Y(4260) $ region of the
  $ \chi_{c1} \, \gamma $ distribution, and six entries
  altogether, about two or three times the number estimated
  as non-$ J/\psi $ $ \mu^- \mu^+ \gamma \gamma $.
  This leads to the question of other potential sources of
  background.

Excluding $ J/\psi \gamma \gamma $ candidates whose
invariant mass lies within 25 MeV/$c^2$ of the $ \psi({\rm 2S})$,
we examined the remaining $ \gamma \gamma $ invariant mass
distribution and separated the pairs into those
from each of the four candidate categories and
those in which the parent did not satisfy the
criteria for any of the intermediate states.
We use this last set of events to estimate
how many additional background events might be observed in
the $ \chi_{c1} \gamma $ sample in the $ Y(4260) $
mass region.
The statistics of the samples are low, so a precise
estimate is not possible.
A reasonable estimate is that the expected background rate is
between 0.5 and 1.0 additional events.
Thus, we cannot exclude the possibility that both events
observed are background.
Neither can we exclude the possibility that they
are signal.
We therefore calculate the 90\% confidence level
upper limit assuming no background.

We calculate the 90\% confidence level upper limit
for the integrated cross section times branching
fraction (or product of branching fractions) for
each of the four channels considered using 2.3 events
as the Poisson upper limit for channels where no entries
are observed in the $ Y(4260) $ region and 5.3 events
in the case of the $ \chi_{C1 \gamma} $ channel
where 2 entries are observed.
Similarly, we use 5.3 events
for calculating the
upper limit for inclusive $ J/\psi \gamma \gamma $
production of these states.
The Monte Carlo efficiencies for the
$ J/\psi \eta $, $ J/\psi  \pi^0 $,
and $ \chi_C \gamma $ channels are 11.2\%,
10.9\%, and 11.9\% respectively.
The efficiency for inclusive $ J/\psi \gamma \gamma $
production of these states, 12.2\%, is somewhat higher 
than the efficiencies of the exclusive final states because,
in contrast to
the those efficiencies,
it counts as reconstructed signal also those
events generated as one exclusive final
state but reconstructed as another. 
To allow for systematic uncertainties in luminosity
times reconstruction efficiency, 
we round off each of the precisely calculated
values, increasing them by between 10\% and 20\%.
These  results are shown in Table \ref{NumericalSummaryTable}.
The ratio of branching fractions, with respect to
$ Y(4260) \to J/\psi \pi^- \pi^+ $, is calculated using
the cross section times branching fraction result
from our original study \cite{shuwei}:
$ \Sigma \times {\cal B} (  Y(4260) \to J/\psi \pi^- \pi^+ ))
 = (7.0 \pm 1.3 \pm 1.0) $~MeV~nb.

\begin{table}[tbh]
\label{NumericalSummaryTable}
\begin{center}
\caption{
This table compares the integrated cross section
times branching fraction results for the
$ Y(4260) \to J/\psi \gamma \gamma $ decay modes studied.
Limits relative to $  Y(4260) \to J/\psi \pi^- \pi^+ $
are calculated using the cross section measured
in our original study \cite{shuwei}.
}
\begin{tabular} {|c|c|c|}
\hline\hline
channel $(X) $& $ \Sigma \times {\cal B} (X) $ & $  {\cal B} (X) / 
 {\cal B} ( J/\psi \pi^- \pi^+) $ \\ [8pt] 
        &            &                 \\
$ Y(4260) \to J/\psi \eta $ &
  $ < 10  $  MeV nb  & $ < 1.4 $            \\ [8pt] \hline
$ Y(4260) \to J/\psi \pi^0 $ &
  $ < 4   $   MeV nb     &  $ < 0.6 $         \\ [8pt] \hline
$ Y(4260) \to \chi_{c1} \gamma  \to  J/\psi \gamma \gamma$ &
  $ < 25   $ MeV nb     &  $ < 3.6  $      \\ [8pt] \hline
$ Y(4260) \to \chi_{c2} \gamma  \to  J/\psi \gamma \gamma$ &
  $ < 18   $  MeV nb    &  $ < 2.6 $          \\ [8pt] \hline
$ Y(4260) \to J/\psi \gamma \gamma $ &
  $ < 8 $ MeV nb        &  $ < 1.2 $          \\ [8pt] \hline
\end{tabular}
\end{center}
\end{table}

As it is our benchmark mode,
we also study $ J/\psi \pi^- \pi^+ $
at high mass using selection criteria very similar to
those used for $ J/\psi \gamma \gamma $.
We first studied expected background rates.
We defined wide dilepton mass windows above and below
the $ J/\psi $ mass region in both channels and 
used the $ \ell^+ \ell^- \pi^+ \pi^- $ candidates
in those windows as if they were $ J/\psi \pi^- \pi^+ $
candidates to estimate  background rates.
Scaling the widths of the background windows to those
used for $ J/\psi $ candidates, we estimate 8 background
events in the 2.5 GeV/$c^2$ range from 3.75 GeV/$c^2$ to 6.25 GeV/$c^2$,
distributed more or less uniformly over that range
and divided roughly equally between $ J/\psi \to
\mu^+ \mu^- $ and $ J/\psi \to e^+ e^- $ candidates.

\begin{figure}[thb]
\begin{minipage}{2.9in}
 \centering
\mbox{\epsfxsize7.5cm\epsffile{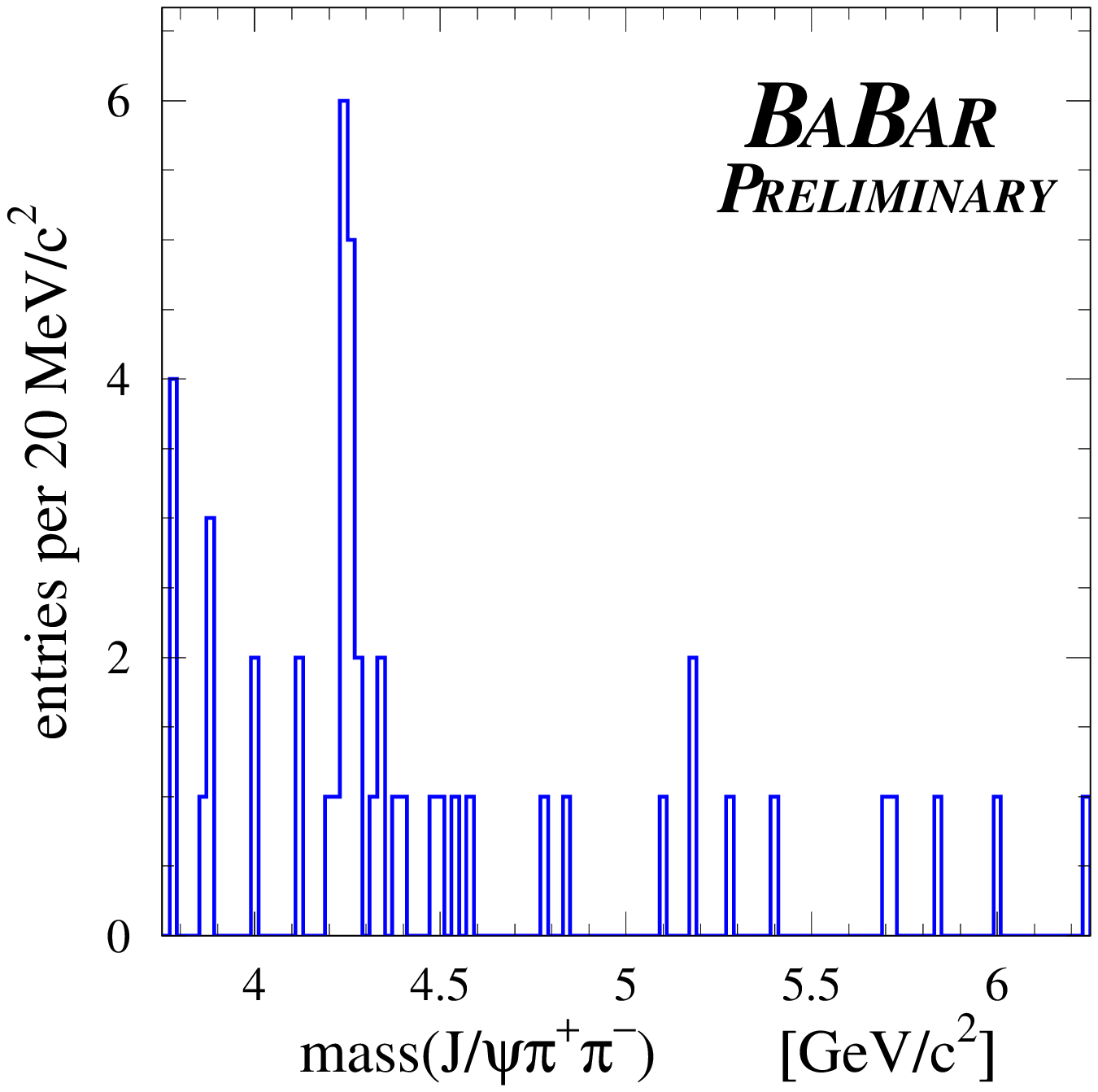}}
\caption{%
This is the
$ J/\psi \, \pi^- \pi^+ $
invariant mass distribution
for candidates satisfying
all final analysis criteria.
}
\label{UnblindedPsiPiPiPlot}
\end{minipage}
\quad
\begin{minipage}{2.9in}
 \centering
\mbox{\epsfxsize7.5cm\epsffile{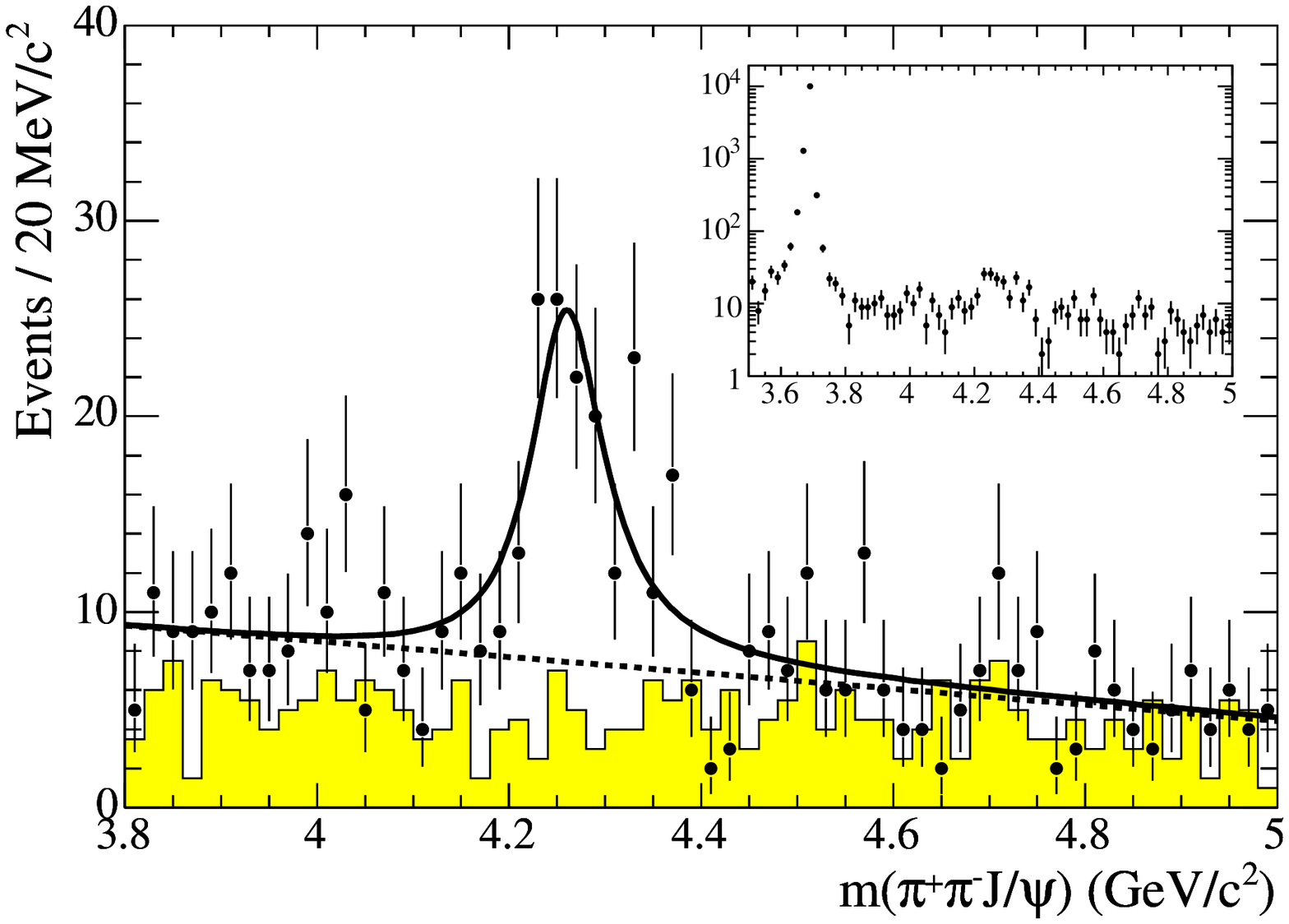}}
\caption{%
This is the $ J/\psi  \pi^- \pi^+ $
invariant mass distribution
presented in the original
$ Y(4260) $ paper \cite{shuwei}.
}
\label{ShuweiPlot}
\end{minipage}
\end{figure}

The high mass  $ J/\psi \pi^- \pi^+ $
sample for this study is shown in Fig.~\ref{UnblindedPsiPiPiPlot}.
For comparison, the plot from 
our original $ Y(4260) $ paper \cite{shuwei} is shown in
Fig.~\ref{ShuweiPlot}.
As anticipated, the signal level is much lower 
in this study
and the signal-to-background
ratio is much higher,
primarily because here we 
require that the ISR photon be detected directly
and not 
produced outside the
EMC acceptance.
The non-$ J/\psi $ $ \mu^- \mu^+ \pi^- \pi^+ $
background rate in this plot,  
estimated from the $ J/\psi $
sidebands,
is 
$ \approx 0.064 $ events per 20 MeV/$c^2$ bin.
It is insufficient to explain the 10 entries
observed
in the  mass range 5.0 GeV/$c^2$ to 6.25 GeV/$c^2$,
where it predicts only 4.
In the window $ \pm 250$~MeV/$c^2$ around
4.260 GeV/$c^2$
we observe 24 events where our background
prediction is 1.6.
The data are consistent with a structure 90 MeV/$ c^2 $
in width, but the statistics are too low to draw an
independent conclusion.
In the original discovery of the $ Y(4260) $ \cite{shuwei},
the 
resonant cross section times branching ratio was measured
to be
$ (7.0 \pm 1.3 \pm 1.0 ) $~MeV~nb.
This predicts a central value for the 
resonant signal level in
the current analysis to be $ \approx 10.5 \pm 2.5 $
events.
The higher event rates observed in the $ Y(4260) $
mass region and at higher mass suggests the possibility of
continuum $ J/\psi \pi^- \pi^+ $
production.

\vspace{-0.2in}
\section{SUMMARY}
\label{sec:Summary}

We observe ISR production of $ \psi({\rm 2S}) $
decaying into
$  J/\psi \eta $, $ \chi_{c1} \gamma $, and
$ \chi_{c2} \gamma $
with summed branching fraction generally
consistent with, but somewhat greater than,
that reported by CLEO.
  We see no events in the $ Y(4260) $ region in the $ J/\psi \, \eta $,
  $ J/\psi  \pi^0 $, or $ \chi_{c2}\, \gamma  $ distributions.
   We observe two entries in the $ Y(4260) $ region of the
  $ \chi_{c1} \, \gamma $ distribution, which is consistent
  with being either signal or background.
We set 90\% confidence level upper limits on the
integrated cross section times branching fraction
(or product of branching fractions) for each of these
channels.
  
We observe ISR production of $ J/\psi \pi^- \pi^+ $
events near 4260 MeV/$c^2$ consistent with the mass
and width originally reported  for the $ Y(4260) $ \cite{shuwei}.
With the improved signal-to-background ratio of the analysis
reported here, we observe an excess of events,
both in the $ Y(4260) $ mass region and
at higher mass.
These observations suggest continuum production of
$ J/\psi \pi^- \pi^+ $.

\section{ACKNOWLEDGMENTS}
\label{sec:Acknowledgments}

We are grateful for the 
extraordinary contributions of our \pep2\ colleagues in
achieving the excellent luminosity and machine conditions
that have made this work possible.
The success of this project also relies critically on the 
expertise and dedication of the computing organizations that 
support \babar.
The collaborating institutions wish to thank 
SLAC for its support and the kind hospitality extended to them. 
This work is supported by the
US Department of Energy
and National Science Foundation, the
Natural Sciences and Engineering Research Council (Canada),
Institute of High Energy Physics (China), the
Commissariat \`a l'Energie Atomique and
Institut National de Physique Nucl\'eaire et de Physique des Particules
(France), the
Bundesministerium f\"ur Bildung und Forschung and
Deutsche Forschungsgemeinschaft
(Germany), the
Istituto Nazionale di Fisica Nucleare (Italy),
the Foundation for Fundamental Research on Matter (The Netherlands),
the Research Council of Norway, the
Ministry of Science and Technology of the Russian Federation, 
Ministerio de Educaci\'on y Ciencia (Spain), and the
Particle Physics and Astronomy Research Council (United Kingdom). 
Individuals have received support from 
the Marie-Curie IEF program (European Union) and
the A. P. Sloan Foundation.

\end{document}